\documentclass[%
preprint,
 amsmath,amssymb,
 aps,
 showkeys,
]{revtex4-1}

\usepackage{graphicx}
\usepackage{dcolumn}
\usepackage{bm}
\usepackage{hyperref}
\usepackage[mathlines]{lineno}
\usepackage{txfonts}

\def\eqref#1{(\ref{#1})}

\begin{document}

\preprint{APS/123-QED}
\title{A comparison of X-ray stress measurement methods \\based on the fundamental equation}

\author{Toshiyuki Miyazaki}
 \email{trmiyazaki@staff.kanazawa-u.ac.jp}
\author{Toshihiko Sasaki}%
\affiliation{%
Kanazawa University, \\Kakuma-machi, Kanazawa, 920-1192, Ishikawa, Japan}%




\date{\today}

\begin{abstract}
Stress measurement methods using X-ray diffraction (XRD methods) are based on so-called fundamental equations.
The fundamental equation is described in the coordinate system that best suites the measurement situation, and, thus, making a comparison between different XRD methods is not straightforward.
However, by using the diffraction vector representation, the fundamental equations of different methods become identical.
Furthermore, the differences between the various XRD methods are in the choice of diffraction vectors and the way of calculating the stress from the measured data.
The stress calculation methods can also be unified using the general least-squares method, which is a common least-squares method of multivariate analysis.
Thus, the only difference between these methods turns out to be in the choice of the set of diffraction vectors.
In light of these ideas, we compare three commonly used XRD methods: the $\sin^2 \psi$ method, the $\text{XRD}^2$ method, and the $\cos \alpha$ method using the estimation of the measurement errors.
\end{abstract}

\pacs{Valid PACS appear here}
                              
\keywords{X-ray stress measurement; Debye--Scherrer ring;  X-ray diffraction}
\maketitle


\section{Introduction}
The $\cos \alpha$ method \cite{Taira78}, an X-ray diffraction (XRD) method, is  widely used in industry, but there are few studies comparing it with other XRD methods in their theoretical aspects.
Although we gave a mathematical explanation of the methods based on Fourier series for the plane stress (biaxial stress) case \cite{Miyazaki14}, it is important to place the $\cos \alpha$ method and the other XRD methods on a common mathematical basis.
In this study, we compare the $\cos \alpha$ method for the triaxial stress case \cite{Sasaki09} with the $\sin^2 \psi$ method (for example, please see \cite{Welzel05}) and the $\text{XRD}^2$ method \cite{He97} from the aspect of the fundamental equation.
First, we show that all three methods are based on a common fundamental equation in the diffraction vector representation.
Second, we show that this fundamental equation can be solved in a common way by using the general least squares method \cite{Winholtz88}.
Accordingly, the only difference between XRD methods is the choice of the set of diffraction vectors.
Finally, we compare XRD methods based on the measurement error estimation.

\section{Fundamental equation}
For the sake of simplicity, we will suppose that the specimen is a polycrystal composed of elastically isotropic crystallites.
Furthermore, we will assume that the microscopic stress of the specimen can be ignored.

We will use the conventional coordinate system of the $\sin^2 \psi$ method (for example, see Fig.~2 of \cite{Welzel05}).
The unit diffraction vector ${\bm n}$ (in the following, we call the unit diffraction vector the ``diffraction vector'') can be described by two angles: $\phi$ and $\psi$ (Fig.~1). $\phi$ is the rotation angle of the diffraction vector around the $S_3$ axis, and $\psi$ describes the tilt angle of the diffraction vector from the $S_3$ axis.
Though \cite{Welzel05} describes the strain corresponding to this diffraction vector as $\varepsilon_{\phi \psi}^{hkl}$, we will consider diffraction by a single diffraction plane ($hkl$) and use $\varepsilon_{\phi \psi}$ for simplicity.
The X-ray measured strain can be described using the strain in the specimen frame of reference as
\begin{align}
	\varepsilon_{\phi \psi} &= \varepsilon_{11} \cos^2 \phi \sin^2 \psi + \varepsilon_{22} \sin^2 \phi \sin^2 \psi + \varepsilon_{33} \cos^2 \psi \notag \\
		&\quad + \varepsilon_{12} \sin (2 \phi) \sin^2 \psi + \varepsilon_{13} \cos \phi \sin (2 \psi) \notag \\
		&\quad + \varepsilon_{23} \sin \phi \sin (2 \psi)
		\label{eq:epsilon01}
\end{align}
This is the fundamental equation of the $\sin^2 \psi$ method (for example, Eq. (13) of \cite{Welzel05}).
Accordingly, the $\sin^2 \psi$ method can be considered an inverse problem of estimating $\varepsilon_{ij}\, (i,j=1\cdots 3)$ from $\varepsilon_{\phi \psi}$ measured with a certain set of $(\phi, \psi)$.
\begin{figure}[hbt]
	\resizebox{\textwidth}{!}{
		\includegraphics{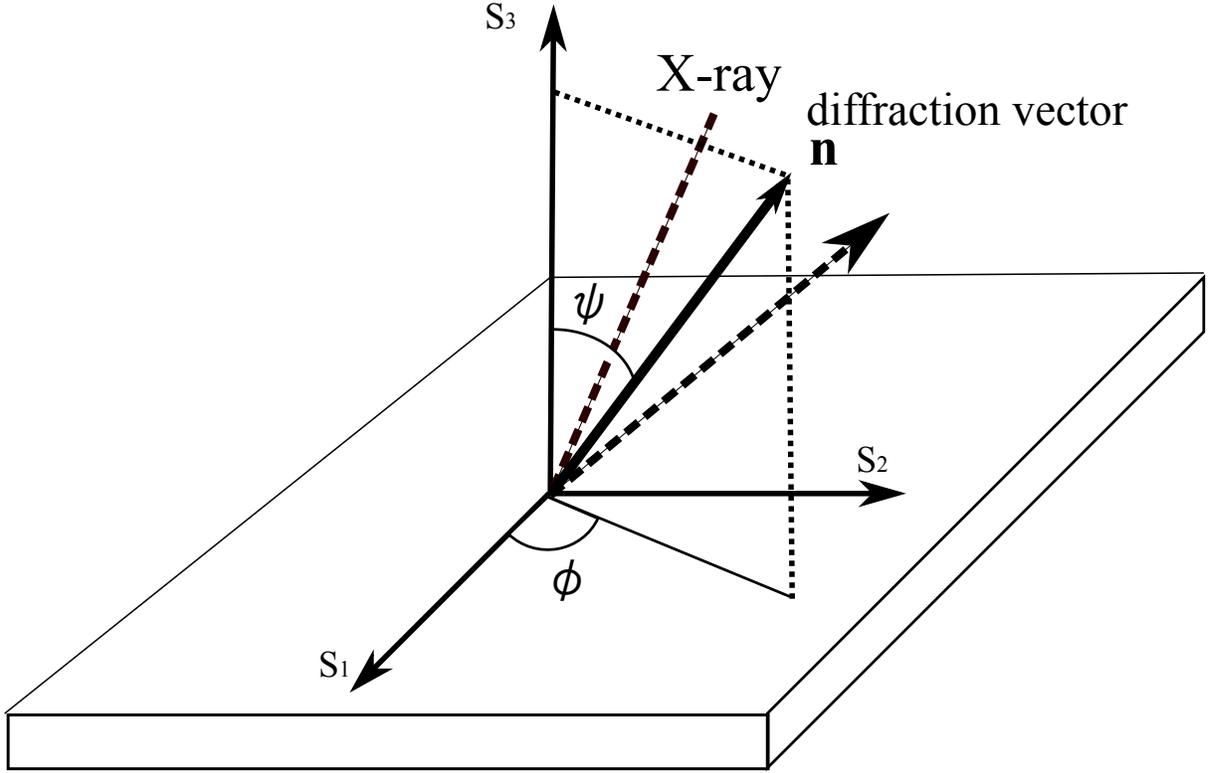}
	}
	\caption{Arrangement of X-ray stress measurement.}
\end{figure}%

The diffraction vector ${\bm n}$ can be described using $(\phi, \psi)$:
\begin{equation}
	{\bm n} =  \left(
		\begin{array}{@{\,}c@{\,}}
			n_1 \\
			n_2 \\
			n_3
		\end{array}
	\right)
	= \left(
		\begin{array}{@{\,}c@{\,}}
			\sin \psi \cos \phi \\
			\sin \psi \sin \phi \\
			\cos \psi
		\end{array}
	\right)
	\label{eq:n01}
\end{equation}

Substituting Eq. \eqref{eq:n01}, Eq. \eqref{eq:epsilon01} becomes
\begin{equation}
	\varepsilon_{\bm n} = n_1^2 \varepsilon_{11} + n_2^2 \varepsilon_{22} + n_3^2 \varepsilon_{33} + 2 n_1 n_2 \varepsilon_{12} +2  n_1 n_3 \varepsilon_{13} +2  n_2 n_3 \varepsilon_{23}
	\label{eq:epsilon02}
\end{equation}
This is the fundamental equation in the diffraction vector representation.

As in the case of Eq. \eqref{eq:n01}, a diffraction vector ${\bm n}$ can be represented by two circumference angles.
Here, we will represent ${\bm n}$ by a $(\phi, \psi)$ pair, which is equivalent to the $(\phi, \psi)$ pair of the $\sin^2 \psi$ method (Fig. 1).
Using $(\phi, \psi)$ pairs, the diffraction vector can be displayed in a pole figure.
Figure~2a shows the definition of the pole figure (angles are in radians).
This figure displays a diffraction vector as a $(\phi, \psi)$ pair and shows the set of diffraction vectors as a constellation.
The center of the figure is $(\phi=0, \psi=0)$, and $\phi$ is the circumference angle.
The distance from the center describes $\sin \psi$.

Figure~2b shows an example of a diffraction vector of the $\sin^2 \psi$ method $(\phi=0, \psi=45^\circ)$.
It has to be emphasized that when measuring $\varepsilon_{\bm n}$ for a $(\phi, \psi)$ pair with a position-insensitive X-ray detector, which many instruments of the $\sin^2 \psi$ method use, several irradiations and detections are required in order to find the peak position of the diffraction ring.
On the other hand, XRD instruments with an area detector require only one X-ray irradiation and detection to find the peak position.
\begin{figure*}[hbt]
	\resizebox{\textwidth}{!}{
		\includegraphics{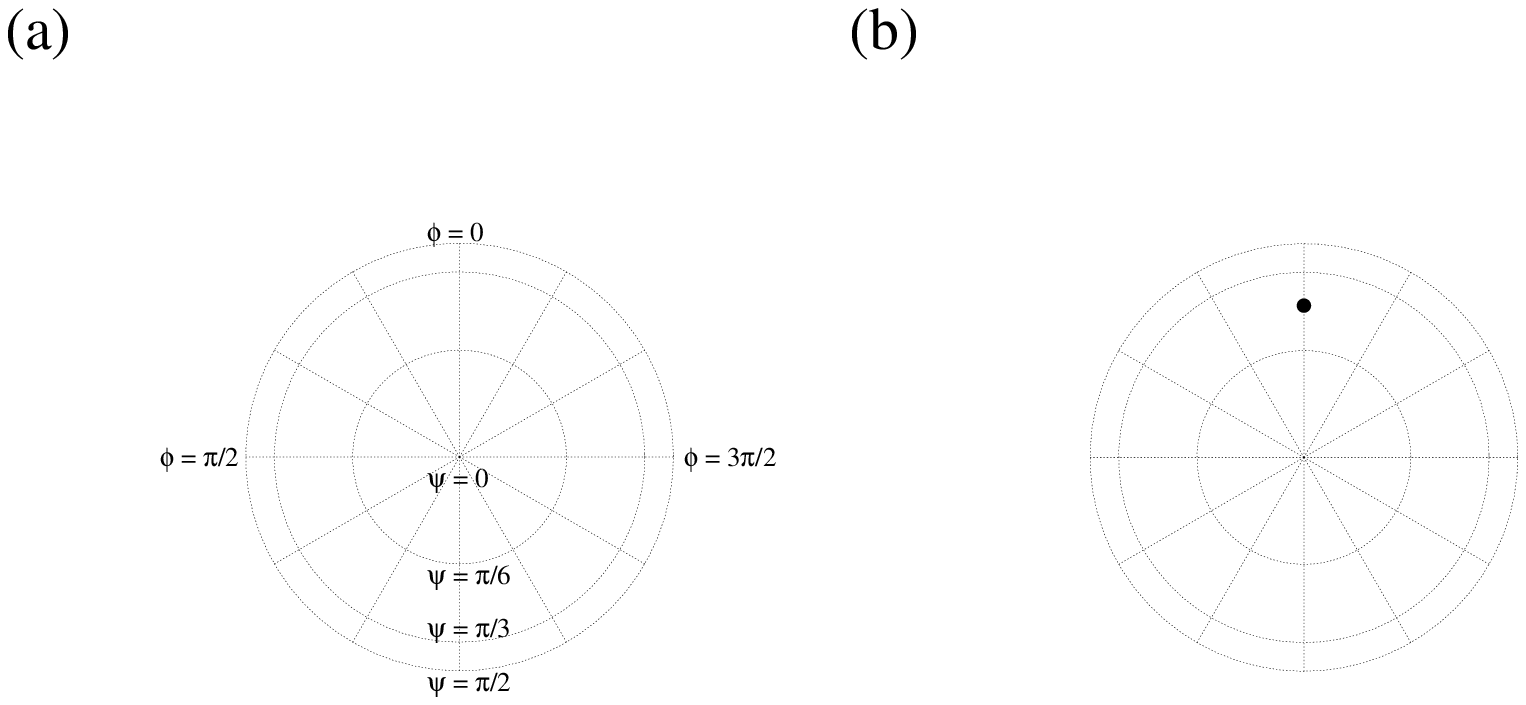}
	}
	\caption{(a) Definition of the pole figure (angles are shown in radians). (b) Example of a diffraction vector of an X-ray irradiation of the $\sin^2 \psi$ method corresponding to $(\phi=0, \psi=45^\circ)$.}
\end{figure*}%

\subsection{$\cos \alpha$ method}
The $\cos \alpha$ method measures the stress from one or more Debye--Scherrer (D--S) rings.
Figure~2 in \cite{Taira78}, Fig.~1 in \cite{Sasaki95a} or Fig.~1 in \cite{Miyazaki14} illustrate the set up of this method.
The diffraction vector of the $\cos \alpha$ method ( Eq. (5) of \cite{Taira78}) is
\begin{equation}
	{\bm n} =  \left(
		\begin{array}{@{\,}c@{\,}}
			n_1 \\
			n_2 \\
			n_3
		\end{array}
	\right)
	= \left(
		\begin{array}{@{\,}c@{\,}}
			\cos \eta \sin \psi_0 \cos \phi_0 - \sin \eta \cos \psi_0 \cos \phi_0 \cos \alpha - \sin \eta \sin \phi_0 \sin \alpha\\
			\cos \eta \sin \psi_0 \sin \phi_0 - \sin \eta \cos \psi_0 \sin \phi_0 \cos \alpha + \sin \eta \cos \phi_0 \sin \alpha\\
			\cos \eta \cos \psi_0 + \sin \eta \sin \psi_0 \cos \alpha
		\end{array}
	\right)
	\label{eq:n02}
\end{equation}
Note that $\psi_0$ of this method is not identical to $\psi$ of Fig. 1.
Using Eq. \eqref{eq:n02}, the fundamental equation of the $\cos \alpha$ method (for example, Eq. (8) of \cite{Taira78}) becomes identical to Eq. \eqref{eq:epsilon02}.

The diffraction vector of Eq. \eqref{eq:n02} can be expressed as an equivalent $(\phi, \psi)$ pair of the $\sin^2 \psi$ method, as follows:
\begin{equation}
	\psi = \cos^{-1} n_3
	\label{eq:psi01}
\end{equation}
and
\begin{equation}
	\phi = 
	\begin{cases}
		\tan^{-1} (n_2/n_1) \,  &(\text{if}\quad n_2 \geq 0)\\
		\tan^{-1} (n_2/n_1) + \pi \, &(\text{if}\quad n_2 < 0)
	\end{cases}
	\label{eq:phi01}
\end{equation}
Figure~3a shows an example pole figure of a constellation of diffraction vectors resulting from an X-ray irradiation.
The conditions of the figure are taken from \cite{Miyazaki14}: $2 \theta = 157.08^\circ$ ($\eta=11.46^\circ$), $\phi_0=0^\circ$, $\psi_0=35^\circ$, and $0^\circ \leq \alpha < 360^\circ$.
It has to be emphasized that this constellation corresponds to a single X-ray irradiation.
Because the $\cos \alpha$ method utilizes the data from a whole D--S ring, it is possible to measure the biaxial stress with a single X-ray irradiation.
To measure the triaxial stress, the $\cos \alpha$ method requires a number of X-ray irradiations with two to four $(\phi_0, \psi_0)$ pairs \cite{Sasaki09, Sasaki95b}.
\begin{figure*}[hbt]
	\resizebox{\textwidth}{!}{
		\includegraphics{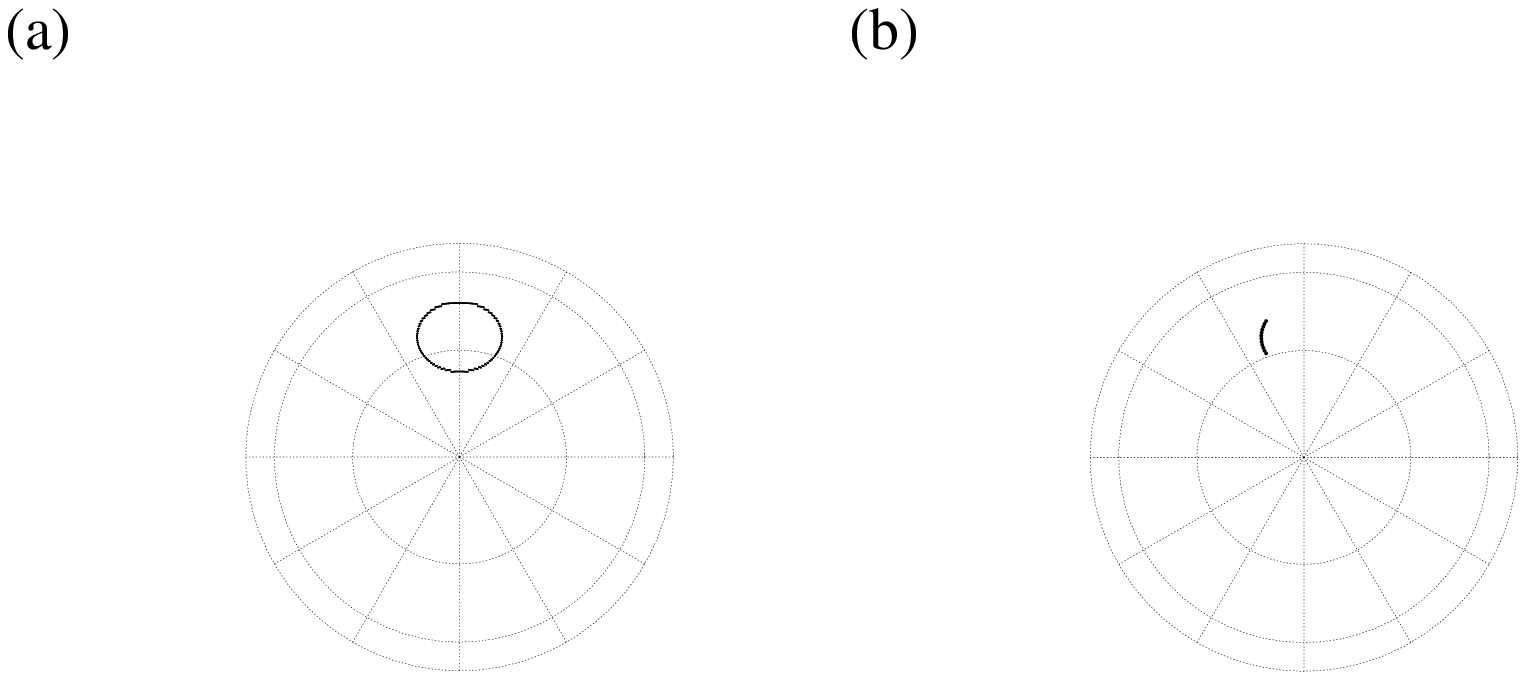}
	}
	\caption{(a) Example pole figure of the diffraction vector of the $\cos \alpha$ method for an X-ray irradiation. (b) Example pole figure of the diffraction vector of the $\text{XRD}^2$ method For an X-ray irradiation.}
\end{figure*}%

\subsection{$\text{XRD}^2$ method}
The $\text{XRD}^2$ method \cite{He97} measures the stress from fractions of D--S rings.
To discuss this method, we will use the coordinate system depicted in Figs. 6 and 10 of \cite{He00}.
The diffraction vector ${\bm n} = (n_1, n_2, n_3)^{\rm T}$ of the method (Eq. (5) in \cite{He00} and shown as $(h_1, h_2, h_3)$) is
\begin{align}
	n_1 &= \sin \theta (\sin \phi \sin \psi \sin \omega + \cos \phi \cos \omega) + \cos \theta \cos \gamma \sin \phi \cos \psi \notag \\
	    &\quad - \cos \theta \sin \gamma (\sin \phi \sin \psi \cos \omega - \cos \phi \sin \omega ) \notag \\
	n_2 &= -\sin \theta (\cos \phi \sin \psi \sin \omega - \sin \phi \cos \omega) - \cos \theta \cos \gamma \cos \phi \cos \psi \notag \\
	    &\quad + \cos \theta \sin \gamma (\cos \phi \sin \psi \cos \omega + \sin \phi \sin \omega ) \notag \\
	n_3 &= \sin \theta \cos \psi \sin \omega - \cos \theta \sin \gamma \cos \psi \cos \omega - \cos \theta \cos \gamma \sin \psi
	\label{eq:n03}
\end{align}
The fundamental equation of the $\text{XRD}^2$ method in the diffraction vector representation is identical to Eq. \eqref{eq:epsilon02}.

The diffraction vector of Eq. \eqref{eq:n03} can be expressed as an equivalent $(\phi, \psi)$ pair of the $\sin^2 \psi$ method using Eqs. \eqref{eq:psi01} and \eqref{eq:phi01}.
Figure~3b shows an example of a constellation of diffraction vectors of the $\text{XRD}^2$ method resulting from an X-ray irradiation.
To make the difference from the $\cos \alpha$ method clear, the angles are: $2 \theta = 157.08^\circ$, $\phi_0 = 90^\circ$, $\psi=35^\circ$ (this $\psi$ is not identical to that of Fig. 1), $\omega=90^\circ$, and $62.5^\circ \leq \gamma \leq 117.5^\circ$.
The range of $\gamma$ was taken from \cite{He03}. Comparing Figs. 3a and 3b, it can be seen that the constellation of diffraction vectors of the $\text{XRD}^2$ is part of that of the $\cos \alpha$ method.
Thus, the $\cos \alpha$ can measure the stress by using less X-ray radiation than that of the $\text{XRD}^2$ method.

\subsection{Comparisons of diffraction vector formulas}
So far, we have seen that the fundamental equations of the $\sin^2 \psi$ method, the $\cos \alpha$ method, and the $\text{XRD}^2$ method are identical in the diffraction vector representation.
In this section, we demonstrate that the expressions of the diffraction vectors (i.e. Eqs. \eqref{eq:n01}, \eqref{eq:n02}, and \eqref{eq:n03}) agree each other with a proper coordinate transformations.
First, we show that Eq. \eqref{eq:n01} is a special case of Eq. \eqref{eq:n02}.
Then we show that Eq. \eqref{eq:n02} is a special case of Eq. \eqref{eq:n03}.

$\sin^2 \psi$ can be regarded as a method that measures only one point on a D--S ring: $\alpha=0$ of the $\cos \alpha$ method.
Thus, substituting $n_1$ of Eq. \eqref{eq:n01} with $\alpha=0$ and $\phi_0 = \phi$, we obtain
\begin{align*}
	n_1 &= \cos \eta \sin \psi_0 \cos \phi_0 - \sin \eta \cos \psi_0 \cos \phi_0 \cos \alpha - \sin \eta \sin \phi_0 \sin \alpha \\
		&= \cos \phi (\sin \psi_0 \cos \eta - \cos \psi_0 \sin \eta) \\
		&= \cos \phi \sin (\psi_0 - \eta) 
\end{align*}
Using $\psi_0 = \psi + \eta$ (Fig. 2 of \cite{Taira78}), we obtain
\[
	n_1 = \cos \phi \sin \psi
\]
With the same substitutions: $\alpha=0$, $\phi_0 = \phi$, and $\psi_0 = \psi + \eta$, we find that Eq. \eqref{eq:n02} is equivalent to Eq. \eqref{eq:n01}.
Thus, the representation of the diffraction vector in the $\sin^2 \psi$ method is a special case of that of the $\cos \alpha$ method.

Comparing the arrangement of the $\text{XRD}^2$ method with the arrangement of the $\cos \alpha$ method, we find that $\gamma = \pi - \alpha$ and $\eta$ satisfies $\theta = \pi/2 - \eta$.
Thus, $n_1$ of Eq. \eqref{eq:n03} can be modified as
\begin{align*}
	n_1 &= \cos \eta ( \sin \phi \sin \psi \sin \omega + \cos \phi \cos \omega) - \sin \eta \cos \alpha \sin \phi \cos \psi \\
	    &\quad - \sin \eta \sin \alpha ( \sin \phi \sin \psi \cos \omega - \cos \phi \sin \omega )
\end{align*}
Furthermore, by setting $\omega =\pi/2$, $\phi = \phi_0 + \pi/2$, and $\psi = \psi_0$, we obtain
\[
	n_1 = \cos \eta \sin \psi_0 \cos \phi_0 - \sin \eta \cos \psi_0 \cos \phi_0 \cos \alpha
			 - \sin \eta \sin \phi_0 \sin \alpha
\]
which is identical to $n_1$ of Eq. \eqref{eq:n02}.
In the similar manner, Eq. \eqref{eq:n03} becomes identical to Eq. \eqref{eq:n02} with the conversions:
\[
	\begin{cases}
	\gamma = \pi - \alpha \\
	\theta = \pi/2 - \eta \\
	\omega = \pi/2 \\
	\phi   = \phi_0 + \pi/2 \\
	\psi   = \psi_0
	\end{cases}
\]
Thus, the representation of the diffraction vector of the $\cos \alpha$ method is a special case of that of the $\text{XRD}^2$ method.

\section{Generalized stress determination}
XRD methods can be regarded as inverse problems to obtain the strain of the specimen as the coefficients of Eq. \eqref{eq:epsilon02} for a certain set of diffraction vectors.
In the strict sense, the $\sin^2 \psi$ method and the $\cos \alpha$ method solve the fundamental equation by using simplified analyses that Ortner named ``linear-regression methods'' \cite{Ortner08}.
Though linear-regression methods are useful when computational power is limited, they are not proper least-squares methods.
The generalized least-squares methods of multivariate analysis, which directly solve Eq. \eqref{eq:epsilon02}, have been discussed by \cite{Winholtz88, He03, Ortner09}.
To make a simple comparison of the methods, we solely use the generalized analysis in the following.
As \cite{Haase14} called the $\sin^2 \psi$ method with the general least-squares method analysis the ``generalized $\sin^2 \psi$ method'', we will call the $\cos \alpha$ method with the general least-squares method analysis the ``generalized $\cos \alpha$ method''.
We will not discuss the difference between the linear-regression methods and the generalized least-squares methods any further.

Let us consider the case of observing $\varepsilon_{\bm n}$ with a set of $k$ diffraction vectors.
${\bm n}_{i} \equiv (n_{i1}, n_{i2}, n_{i3})^{\rm T}$ describes the $i$-th diffraction vector, and corresponding equivalent $(\phi_i, \psi_i)$ pairs can be calculated using Eqs. \eqref{eq:psi01} and \eqref{eq:phi01}.
From Eq. \eqref{eq:epsilon02}, $\varepsilon_{\bm n}$ for ${\bm n}_{i}$ satisfies
\begin{equation}
	\varepsilon_{{\bm n}_i} = n_{i1}^2 \varepsilon_{11} + n_{i2}^2 \varepsilon_{22} + n_{i3}^2 \varepsilon_{33} 
				+ 2 n_{i1} n_{i2} \varepsilon_{12} + 2 n_{i1} n_{i3} \varepsilon_{13} + 2 n_{i2} n_{i3} \varepsilon_{23}
	\label{eq:epsilon11}
\end{equation}

Let us define a $k \times 6$ matrix embodying the coefficients of Eq. \eqref{eq:epsilon11}:
\begin{equation}
	{\bm F} \equiv \left(
		\begin{array}{@{\,}cccccc@{\,}}
			n_{11}^2 & n_{12}^2 & n_{13}^2 & 2 n_{11} n_{12} & 2 n_{11} n_{13} & 2 n_{12} n_{13} \\
			n_{21}^2 & n_{22}^2 & n_{23}^2 & 2 n_{21} n_{22} & 2 n_{21} n_{23} & 2 n_{22} n_{23} \\
			\vdots   & \vdots   & \vdots   & \vdots          & \vdots          & \vdots          \\
			n_{k1}^2 & n_{k2}^2 & n_{k3}^2 & 2 n_{k1} n_{k2} & 2 n_{k1} n_{k3} & 2 n_{k2} n_{k3}
		\end{array}
	\right)
	\label{eq:F01}
\end{equation}
Using this matrix, the set of fundamental equations can be described as
\begin{equation}
	{\boldsymbol \varepsilon_{\bm n}} \equiv
	\left(
		\begin{array}{@{\,}c@{\,}}
			\varepsilon_{{\bm n}_1} \\
			\varepsilon_{{\bm n}_2} \\
			\vdots \\
			\varepsilon_{{\bm n}_k}
		\end{array}
	\right) = {\bm F} \left(
		\begin{array}{@{\,}c@{\,}}
			\varepsilon_{11} \\
			\varepsilon_{22} \\
			\varepsilon_{33} \\
			\varepsilon_{12} \\
			\varepsilon_{13} \\
			\varepsilon_{23}
		\end{array}
	\right)
	\label{eq:epsilon12}
\end{equation}

The strains $(\varepsilon_{11}, \varepsilon_{22}, \varepsilon_{33}, \varepsilon_{12}, \varepsilon_{13}, \varepsilon_{23})^{\rm T}$ can be related to the stresses ${\boldsymbol \sigma} \equiv (\sigma_{11}, \sigma_{22}, \sigma_{33}, \sigma_{12}, \sigma_{13}, \sigma_{23})^{\rm T}$ as
\begin{equation}
	\left(
		\begin{array}{@{\,}c@{\,}}
			\varepsilon_{11} \\
			\varepsilon_{22} \\
			\varepsilon_{33} \\
			\varepsilon_{12} \\
			\varepsilon_{13} \\
			\varepsilon_{23}
		\end{array}
	\right) = \dfrac{1}{E} \left(
		\begin{array}{@{\,}cccccc@{\,}}
			1    & -\nu & -\nu & 0 & 0& 0 \\
			-\nu &    1 & -\nu & 0 & 0& 0 \\
			-\nu & -\nu &    1 & 0 & 0& 0 \\
			   0 &    0 &    0 & 1 & 0& 0 \\
			   0 &    0 &    0 & 0 & 1& 0 \\
			   0 &    0 &    0 & 0 & 0& 1 
		\end{array}
	\right)
	\left(
		\begin{array}{@{\,}c@{\,}}
			\sigma_{11} \\
			\sigma_{22} \\
			\sigma_{33} \\
			\sigma_{12} \\
			\sigma_{13} \\
			\sigma_{23}
		\end{array}
	\right) \equiv {\bm S} 	\boldsymbol \sigma
	\label{eq:S01}
\end{equation}
where $E$ and $\nu$ are the X-ray Young's modulus and Poisson's ratio, respectively.
Substituting Eq. \eqref{eq:S01}, Eq. \eqref{eq:epsilon12} becomes
\begin{equation}
	{\boldsymbol \varepsilon_{\bm n}} = {\bm F} \cdot {\bm S} \boldsymbol \sigma \equiv {\bm M} \boldsymbol \sigma
	\label{eq:epsilon13}
\end{equation}
The general least-squares solution of Eq. \eqref{eq:epsilon13} is
\begin{equation}
	{\boldsymbol \sigma} = {\bm M}^\dagger {\boldsymbol \varepsilon_{\bm n}}
	\label{eq:sigma11}
\end{equation}
where ${\bm M}^\dagger$ is the Moore--Penrose's general inverse of ${\bm M}$.
Equation \eqref{eq:sigma11} is the universal solution for the XRD measurement.

Using $m_{ij}^\dagger$, the $(i, j)$th component of ${\bm M}^\dagger$, and $\varepsilon_{{\bm n}_i}$, each component of $\boldsymbol \sigma$ can be described as a linear combination.
For example,
\begin{equation*}
	\sigma_{11} = \sum_{i=1}^k m_{1i}^\dagger\, \varepsilon_{{\bm n}_i}
\end{equation*}
If the measurement error of each $\varepsilon_{{\bm n}i}$ is independent and has a deviation $\delta \varepsilon$, the measurement error of $\sigma_{11}$ is
\begin{equation}
	\delta \sigma_{11} = \delta \varepsilon \sqrt{\sum_{i=1}^k \bigl(m_{1i}^\dagger \bigr)^2}
	\label{eq:sigma12}
\end{equation}
The errors of the other components of ${\boldsymbol \sigma}$ can be estimated in a similar way.
However, not all of the measurement points of the $\cos \alpha$ method are independent, and the assumption that errors are independent is not fully satisfied.
In this case, Eq. \eqref{eq:sigma12} underestimates the error and an adequately sparse set of $\varepsilon_{{\bm n}_i}$ is required to estimate the correct error.

\section{Comparison of triaxial stress measurement}
In the previous sections, we described the way to calculate stress from $\varepsilon_{\bm n}$ measured for a set of diffraction vectors ${\bm n}$.
Though the stress can be calculated using Eq. \eqref{eq:sigma11}, the equation does not tell which set of diffraction vectors should be chosen to measure the stress. However, once the set of diffraction vectors is chosen, we can estimate the error of the stress measurement by using Eq. \eqref{eq:sigma12}.
In this section, we compare XRD methods in terms of their errors as estimated by Eq. \eqref{eq:sigma12}.
Specifically, we compared representative constellations of the $\sin^2 \psi$ method and the $\text {XRD}^2$ method, three constellations in \cite{Sasaki09} and a new constellation of the $\cos \alpha$ method.
We assumed that the $hkl = 211$ diffraction plane of an $\alpha$-Fe specimen was measured with $\text{Cr-K}_\alphaup$ characteristic X-rays.
The diffraction angle was taken to be $\theta = 78^\circ$ (i.e. $\eta = 12^\circ$), and X-ray Young's modulus and Poisson's ratio were
\begin{align*}
	E &= 221\quad \text{(GPa)}\\
	\nuup &= 0.28
\end{align*}

Table~1 shows the $(\phi, \psi)$ pairs for the generalized $\sin^2 \psi$ method \cite{Dolle80}.
This set requires 31 $(\phi, \psi)$ pairs.
In the case of the $\sin^2 \psi$ method, this means 31 individual data acquisitions (for convenience, we call them ``frames'' hereafter) are required.
As stated before, when using a stress measurement instrument with a position-insensitive X-ray detector, one frame requires several X-ray irradiations.
The pole figure of this constellation is shown in Fig.~4a.
\begin{table}[htbp]
	\caption{Constellation of $(\phi, \psi)$ pairs for the triaxial stress measurement with the $\sin^2 \psi$ method.}
	\label{table:sin01}
	\begin{center}
		\begin{tabular}{c|c}
		$\phi$	& $\psi$ \\
		\hline \hline
		$0^\circ$ & $0^\circ$, $\pm 18^\circ$, $\pm 26^\circ$, $\pm 33^\circ$, $\pm 39^\circ$, $\pm 45^\circ$ \\
		\hline
		$45^\circ$ & \phantom{$0^\circ$, } $\pm 18^\circ$, $\pm 26^\circ$, $\pm 33^\circ$, $\pm 39^\circ$, $\pm 45^\circ$ \\
		\hline
		$90^\circ$ & \phantom{$0^\circ$, } $\pm 18^\circ$, $\pm 26^\circ$, $\pm 33^\circ$, $\pm 39^\circ$, $\pm 45^\circ$ \\
		\end{tabular}
	\end{center}
\end{table}%
\begin{figure*}[hbt]
	\resizebox{\textwidth}{!}{
		\includegraphics{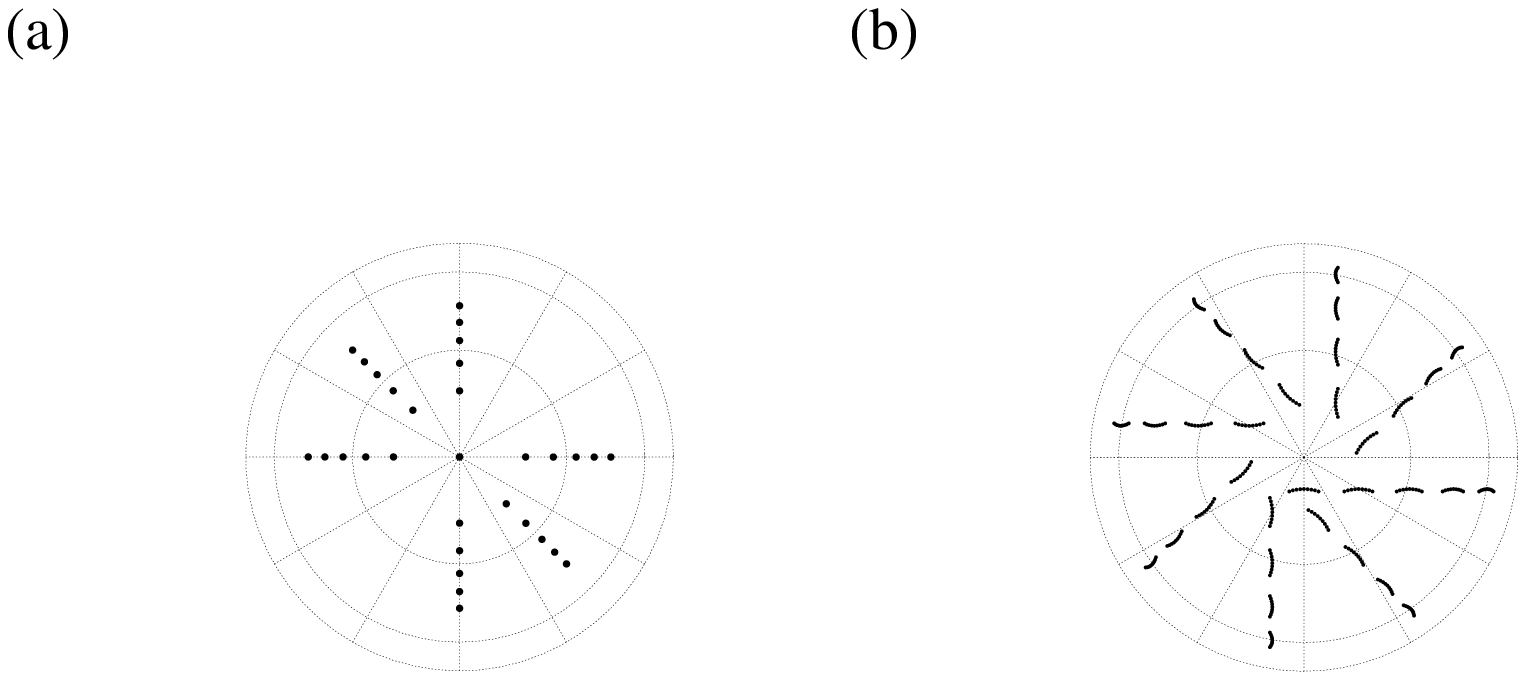}
	}
	\caption{(a) Example pole figure of the diffraction vector of the $\sin^2 \psi$ method for a triaxial stress measurement. (b) Example pole figure of the diffraction vector of the $\text{XRD}^2$ method for a triaxial stress measurement.}
\end{figure*}%

Table~2 shows the pairs of $(\phi, \psi)$ pairs for the $\text{XRD}^2$ method.
Note that the pairs $(\phi, \psi)$ of this table are those of Eq. \eqref{eq:n03} and are not identical to the equivalent $(\phi, \psi)$ pairs of the $\sin^2 \psi$ method.
In the following calculations, we set $\omega = 110^\circ$ and $70^\circ \leq \gamma \leq 110^\circ$ (in $5^\circ$ step) \cite{Takakuwa13}.
This set consists of 33 frames (data acquisitions).
In the case of the $\text{XRD}^2$ method, one frame can be acquired with a single X-ray irradiation.
The pole figure of this constellation is shown in Fig.~4b.
\begin{table}[htbp]
	\caption{Constellation of $(\phi, \psi)$ pairs for the triaxial stress measurement with the ${\text XRD}^2$ method.}
	\label{table:XRD01}
	\begin{center}
		\begin{tabular}{c|c}
		$\phi$	& $\psi$ \\
		\hline \hline
		$0^\circ$ & $0^\circ$ \\
		\hline
		$15^\circ$ & $0^\circ$, $45^\circ$, $90^\circ$, $135^\circ$, $180^\circ$, $225^\circ$, $270^\circ$, $315^\circ$ \\
		\hline
		$30^\circ$ & $0^\circ$, $45^\circ$, $90^\circ$, $135^\circ$, $180^\circ$, $225^\circ$, $270^\circ$, $315^\circ$ \\
		\hline
		$45^\circ$ & $0^\circ$, $45^\circ$, $90^\circ$, $135^\circ$, $180^\circ$, $225^\circ$, $270^\circ$, $315^\circ$ \\
		\hline
		$60^\circ$ & $0^\circ$, $45^\circ$, $90^\circ$, $135^\circ$, $180^\circ$, $225^\circ$, $270^\circ$, $315^\circ$ \\
		\end{tabular}
	\end{center}
\end{table}%
\begin{figure*}[hbt]
	\resizebox{\textwidth}{!}{
		\includegraphics{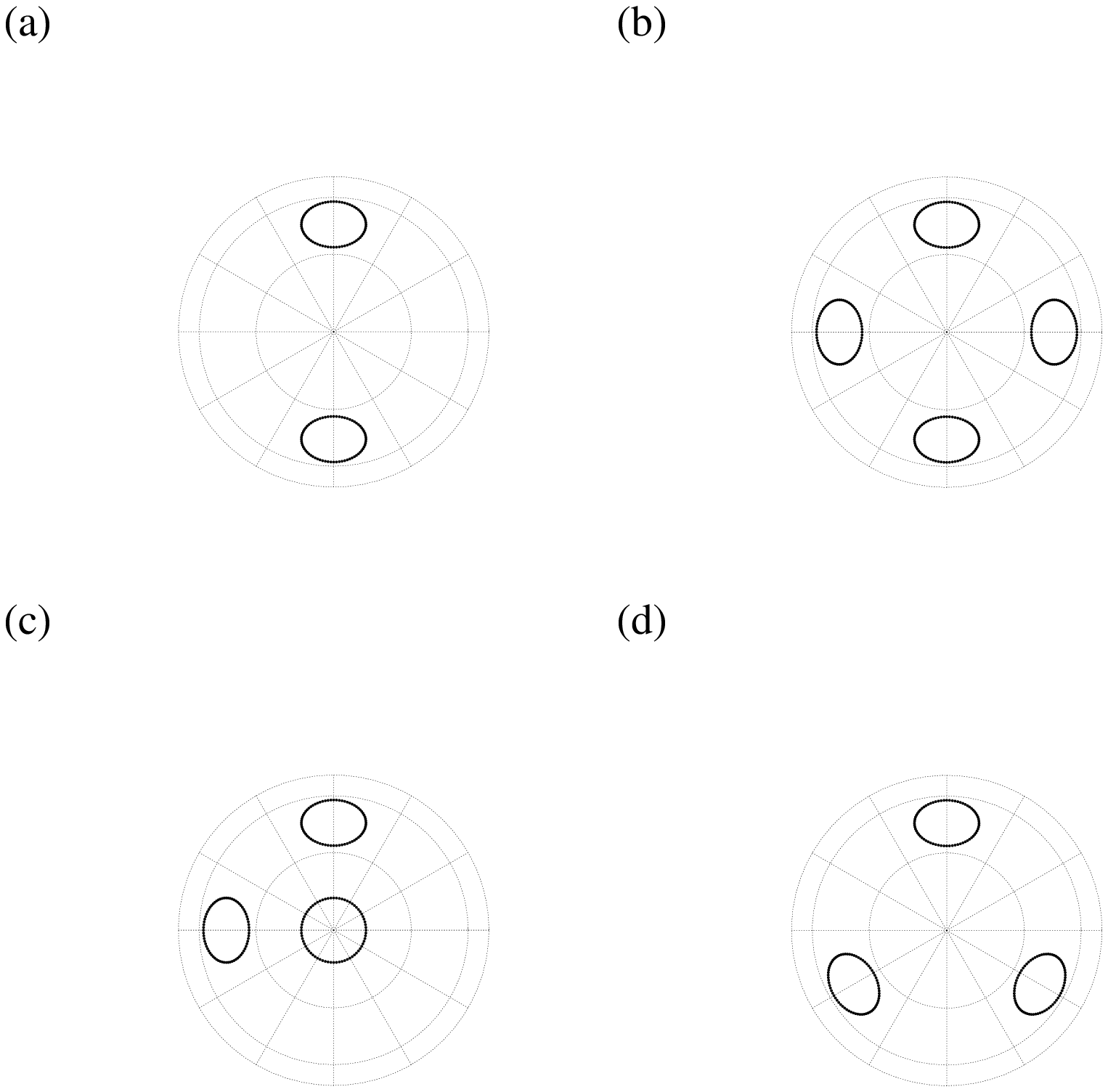}
	}
	\caption{(a) Pole figure of the diffraction vector of the $\cos \alpha$ method for a triaxial stress measurement (Type A). (b) Pole figure of the diffraction vector of the $\cos \alpha$ method for a triaxial stress measurement (Type B). (c) Pole figure of the diffraction vector of the $\cos \alpha$ method for a triaxial stress measurement (Type C). (d) Pole figure of the diffraction vector of the $\cos \alpha$ method for a triaxial stress measurement (Type D).}
\end{figure*}%

Table~3 shows the $(\phi_0, \psi_0)$ pairs for the generalized $\cos \alpha$ method.
Type A is according to \cite{Sasaki95b}, and Types B and C are according to \cite{Sasaki09}.
Type D is new.
Type A requires two frames (data acquisitions), Type B requires four frames, and Types C and D require three frames.
Compared with the other methods, the $\cos \alpha$ method requires the fewest frames.
Figures~5a-d show the pole figures of Type A-D.
\begin{table}[htbp]
	\caption{Constellations of $(\phi_0, \psi_0)$ pairs for the triaxial stress measurement with the generalized $\cos \alpha$ method.}
	\label{table:cosa01}
	\begin{center}
		\begin{tabular}{|c||c|c|}
		\hline
		 Method name & Combinations of ($\phi_0$, $\psi_0$) & No. of Frames \\
		\hline \hline
		Type A & ($0^\circ$, $45^\circ$), ($180^\circ$, $45^\circ$) & 2\\
		\hline
		Type B & ($0^\circ$, $45^\circ$), ($90^\circ$, $45^\circ$), ($180^\circ$, $45^\circ$), ($270^\circ$, $45^\circ$) & 4\\
		\hline
		Type C & ($0^\circ$, $45^\circ$), ($90^\circ$, $45^\circ$), ($0^\circ$, $0^\circ$) & 3\\
		\hline
		Type D & ($0^\circ$, $45^\circ$), ($120^\circ$, $45^\circ$), ($240^\circ$, $45^\circ$) & 3\\
		\hline
		\end{tabular}
	\end{center}
\end{table}%

The generalized $\cos \alpha$ method calculates the stress using whole D--S rings.
The number of data points of one frame is $n_{\alpha}=500$ \cite{Miyazaki14}.
On the other hand, the error of the stress $\deltaup \sigma$ estimated using Eq. \eqref{eq:sigma12} is proportional to $1/\sqrt{n_{\alpha}}$.
From this, one may conclude that the accuracy of the stress measurement can be infinitely improved if $n_{\alpha}$ is increased.
But as stated previously, the neighboring points of a frame are correlated with each other and the effective number of independent data points is less than 500.
Here, we will not discuss the most proper $n_{\alpha}$, but will instead assume $n_{\alpha}=72$ ($5^\circ$ step) in accordance with the $\text{XRD}^2$ method.
This assumption is realistic for the error estimation and sufficient for the purpose of comparison with other methods.

Table~4 shows the error estimated using Eq. \eqref{eq:sigma12}.
Though the values of the $\sin^2 \psi$ method are not identical to those of \cite{Winholtz88}, the differences are small.
The reason for these small discrepancies is under investigation.
The $\text{XRD}^2$ method which consists of 33 frames showed the best accuracy.
Compared with the generalized $\sin^2 \psi$ method, the $\text{XRD}^2$ method is approximately six times more accurate and uses a similar number of frames (31 frames).
\begin{table*}[htbp]
	\caption{Comparisons of the error estimated using Eq. \eqref{eq:sigma12}. $\deltaup \varepsilon=10^{-4}$ and $n_{\alpha}=72$ were assumed.}
	\label{table:error01}
	\begin{center}
		\begin{tabular}{|c||c|c|c|c|c|c|}
		\hline
		 & \multicolumn{6}{c|}{Estimated errors (MPa)}\\
		 \cline{2-7}
		 & $\deltaup \sigma_{11}$ & $\deltaup \sigma_{22}$ & $\deltaup \sigma_{33}$ & $\deltaup \sigma_{12}$ & $\deltaup \sigma_{13}$ & $\deltaup \sigma_{23}$ \\
		\hline \hline
		$\sin^2 \psi$ method  & 36.6 &  36.6 & 15.4 & 26.0 &  7.0 & 7.0  \\
		\hline \hline
		$\text{XRD}^2$ method &  5.6 &   5.6 &  2.8 &  3.6 &  2.0 & 2.0  \\
		\hline \hline
		\multicolumn{1}{|c}{$\cos \alpha$ method} & \multicolumn{6}{c|}{} \\
		\hline 
		Type A                & 50.7 & 138.1 & 50.7 &  8.4 &  1.8 &  8.4 \\
		\hline
		Type B                &  6.2 &   6.2 &  3.1 &  5.9 &  1.8 &  1.8 \\
		\hline
		Type C                & 15.7 &  15.7 &  8.3 &  9.4 &  5.8 &  5.8 \\
		\hline
		Type D                &  8.2 &   8.2 &  3.6 &  6.0 &  3.3 &  3.3 \\
		\hline
		\end{tabular}
	\end{center}
\end{table*}%

The generalized $\cos \alpha$ method showed good accuracy when more than three frames are taken (i.e., Types B--D).
Type B with four frames is as accurate as the $\text{XRD}^2$ method.
This result can be understood intuitively in that a single frame of the $\text{XRD}^2$ method takes 1/8th of the D--S ring, while a single frame of the $\cos \alpha$ method acquires a whole D--S ring.
Thus, the $\cos \alpha$ method can achieve similar accuracy with 1/8th of the frames of the $\text{XRD}^2$ method.
Moreover, by using Type D, we can reduce the number of frames by one while losing only lose 30\% of the accuracy.
Consequently, we recommend Type D for the triaxial stress measurement with the generalized $\cos \alpha$ method.

\section{Summary}
This study showed that the $\sin^2 \psi$, $\cos \alpha$, and $\text{XRD}^2$ methods can be described with a common fundamental equation using the diffraction vector representation.
By fitting the data with the generalized least-squares method, the only differences between these methods are in the choice of the set of diffraction vectors.
The differences between the sets of diffraction vectors become clear in the pole figure plot.
We also estimated the errors of the XRD methods for typical choices of diffraction vector and demonstrated that the $\text{XRD}^2$ method with 33 frames is the most accurate.
We further showed that the generalized $\cos \alpha$ method with four frames is comparable in accuracy to the $\text{XRD}^2$ method.
However, from the viewpoint of the balance between the number of the frames and the accuracy, the generalized $\cos \alpha$ method with three equally spaced frames is recommended.
In the future, the authors will test the conclusions of this study by making actual measurements.

\begin{acknowledgments}
This work was partially supported by a Grant--in--Aid for the Innovative Nuclear Research and Development Program (No.~120804) from the Ministry of Education, Culture, Sports, Science and Technology in Japan.
\end{acknowledgments}


\bibliography{references}

\end{document}